\documentclass[aps,physrev,reprint,superscriptaddress]{revtex4-2}
\usepackage{graphicx}
\usepackage{amsmath}
\usepackage{braket}
\usepackage{hyperref}
\begin{document}

%\preprint{}

\title{Breaking the Moss rule}


\author{Søren Raza}
\email{sraz@dtu.dk}
\affiliation{Department of Physics, Technical University of Denmark, Fysikvej, DK-2800 Kongens Lyngby, Denmark}

\author{Kristian Sommer Thygesen}
\email{thygesen@fysik.dtu.dk}
\affiliation{Department of Physics, Technical University of Denmark, Fysikvej, DK-2800 Kongens Lyngby, Denmark}

\author{Gururaj Naik}
\email{guru@rice.edu}
\affiliation{Electrical \& Computer Engineering, Rice University, 6100 Main Street, Houston, 77005, TX, USA}


\date{\today}

\begin{abstract}
Photonic devices depend critically on the dielectric materials from which they are made, with higher refractive indices and lower absorption losses enabling new functionalities and higher performance. However, these two material properties are intrinsically linked through the empirical Moss rule, which states that the refractive index of a dielectric decreases as its band gap energy increases. Materials that surpass this rule, termed super-Mossian dielectrics, combine large refractive indices with wide optical transparency and are therefore ideal candidates for advanced photonic applications. This Review surveys the expanding landscape of high-index dielectric and semiconductor materials, with a particular focus on those that surpass the Moss rule. We discuss how electronic band structures with a large joint density of states near the band edge give rise to super-Mossian behavior and how first-principles computational screening can accelerate their discovery. Finally, we establish how the refractive index sets the performance limits of nanoresonators, waveguides, and metasurfaces, highlighting super-Mossian dielectrics as a promising route toward the next performance leap in photonic technologies.
\end{abstract}

%\keywords{}

\maketitle


\section{Introduction}
Dielectric and plasmonic materials make up almost all nanophotonic devices. The properties of these materials ultimately limit their functionality and performance. A simple connection between the optical properties of materials and the ultimate performance of devices that employ them can greatly aid design. In addition, such a relationship can significantly benefit material scientists in discovering and engineering better nanophotonic materials. Plasmonic materials have seen a systematic effort in this direction~\cite{blaber2009search,west2010searching,boltasseva2011low}. Plasmonic materials beyond noble metals were identified and evaluated for their performance in various device applications. Such comprehensive studies are lacking for dielectrics. With dielectric nanophotonics gaining prominence because of its smaller losses and ability to confine light to the nanoscale~\cite{Kuznetsov2016}, the need to identify and evaluate various dielectric materials for their performance in nanophotonic devices is crucial~\cite{baranov2017all}. This review presents a case for super-Mossian high-refractive-index dielectrics, reviews recent efforts in this direction, and connects the performance limits of various nanophotonic devices to the refractive index of dielectrics. This work aims to guide material scientists in developing novel nanophotonic dielectrics and aid nanophotonic designers in choosing the right dielectric materials.

Dielectric materials with a high refractive index and low optical absorption are desirable for almost all nanophotonic applications. While a high refractive index enables a tighter confinement of light, low optical absorption guarantees high performance. Hence, materials with a high refractive index and low absorption are desired not only in linear optics, but also in non-linear optics. Many non-linear properties, such as second harmonic generation and the Pockels electro-optic effect, depend directly on the refractive index raised to an exponent of at least two~\cite{Boyd2020}. Silicon is a serendipitous material choice that has paved the way for silicon nanophotonics~\cite{cao2010tuning,lin2014dielectric,Staude2017}. Given the development of silicon-based nanoelectronic device fabrication, silicon is a promising dielectric platform for nanophotonics. However, having only a few materials in a designer's library is a severe limitation. Hence, the question follows: Are there dielectric materials similar to or even better than silicon?
\begin{figure*}[!t]
    \centering
    \includegraphics[width=18cm]{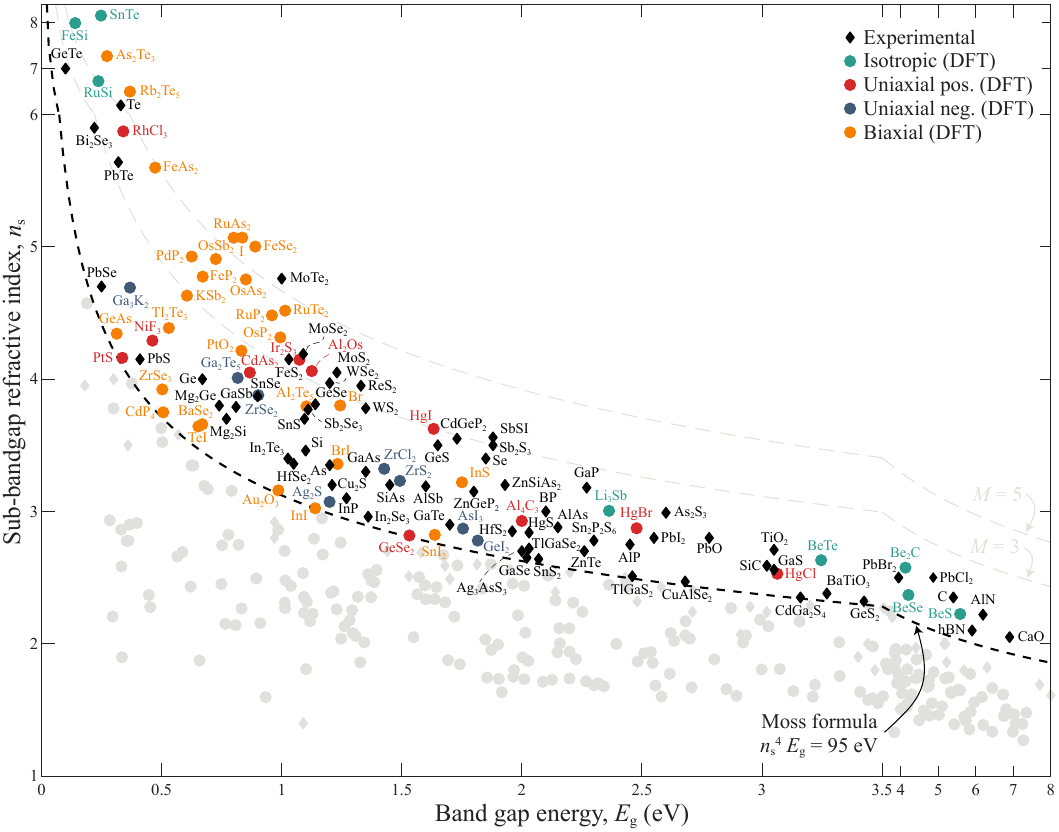}
    \caption{\textbf{Super-Mossian dielectric materials.} Sub-bandgap refractive index $n_\textrm{s}$ as a function of band gap energy $E_\textrm{g}$ for 388 semiconductors, of which 108 are experimental values (see Supplementary~Table~S1) and 280 are calculated using density functional theory (DFT)~\cite{Svendsen2022}. The experimental indices are shown as a function of the fundamental band gap energy, while the calculated indices are as a function of the direct band gap energy. The computed materials are categorized according to their anisotropy. For both experimental and computed materials, the component of the refractive index tensor with the highest value is shown. Frequency-dependent complex refractive indices for all computed materials are available in the CRYSP database~\cite{CRYSP}. The dashed lines show Moss relations with Moss factors of $M=3$ and $M=5$. Note the change in scales on the $E_\textrm{g}$ and $n_\textrm{s}$ axes at 3.5~eV and 6, respectively.}
    \label{fig:figure1}
\end{figure*}

A quest for dielectrics with a high refractive index and low loss has intrigued researchers for decades. Early studies observed that the sub-bandgap refractive index of a semiconductor tends to decrease as its absorption edge or the direct band gap increases. In 1950, Moss quantified this empirical relationship as $E_\textrm{g}n_\textrm{s}^4 = 95$~eV, where $E_\textrm{g}$ is the fundamental band gap energy and $n_\textrm{s}$ is the long wavelength (sub-bandgap) refractive index~\cite{moss1950}. This is commonly known as Moss' rule. Figure~\ref{fig:figure1} displays the refractive index versus band gap for a wide range of materials, including common semiconductors, along with Moss’ rule. While Moss' rule captures the general inverse relation between band gap and refractive index, many materials exceed this prediction. We classify these as super-Mossian materials and define the Moss factor $M$ as 
\begin{equation} \label{eq:MossFactor}
    M = \frac{E_\textrm{g}n_\textrm{s}^4}{95~\textrm{eV}},
\end{equation}
such that $M>1$ corresponds to super-Mossian behavior. Moss himself later noted that materials such as silicon, germanium, and diamond exhibit $M\approx 2$~\cite{moss1951}. Subsequent theoretical work refined this understanding. Finkenrath rigorously derived Moss' rule for semiconductors with zincblende and diamond lattices \cite{finkenrath1964naturf,Finkenrath1988}. while Penn proposed a simple model for isotropic dielectrics~\cite{penn1962wave}. Several other models have since been developed to extend the applicability of these trends to different materials and band gap ranges~\cite{ravindra1979penn,herve1994general,reddy1992analysis,tripathy2015refractive,Gomaa2021}. This long and evolving history has set the foundation for the ongoing search for new super-Mossian materials~\cite{Moss1985,ravindra2007energy,baranov2017all,khurgin2022expanding}.

This paper reviews advances in the recent decade in the pursuit of high-refractive-index semiconductors that surpass the Moss rule. In the following sections, we identify the physical origin of the super-Mossian behavior of semiconductors and review recent significant developments in the search for super-Mossian materials, considering both experimental and computational approaches. Then, we formulate figures of merit to assess the performance of novel dielectric materials in nanophotonic devices and conclude with an outlook.

\section{The origin of super-Mossian behavior}
\begin{figure*}
    \centering
    \includegraphics[width=2\columnwidth]{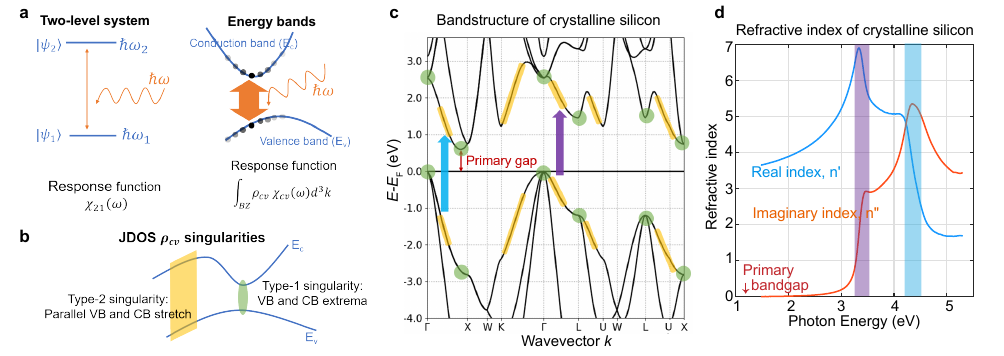}
    \caption{\textbf{Optical properties of semiconductors}. \textbf{a} The optical response of a two-level system under weak excitation is captured by the two-level response function $\chi_{21}$, while the optical response of a semiconductor is the sum of responses from each pair of states. Joint density of states (JDOS) is the total number of state pairs available for optical transitions. \textbf{b} Singularities in JDOS or zeros of \textbf{k}-space gradient of ($E_\textrm{c}-E_\textrm{v}$) shape the optical properties of semiconductors. These singularities occur at the joint extrema of bands (type-1) and parallel stretches of bands (type-2). \textbf{c} Band structure of silicon, highlighting the singularities in JDOS. \textbf{d} Refractive index of silicon relating the prominent features to the singularities in JDOS shown in \textbf{c}.}
    \label{fig:figure2}
\end{figure*}
The refractive index of a semiconductor arises from its electric polarization response. In the weak excitation limit, the linear optical response of a semiconductor can be described by its electrical susceptibility $\chi$. This susceptibility is related to the refractive index $n$ by $n^2=\chi +1$. An expression for $\chi$ may be derived from a semi-classical model of light-matter interaction. Consider a two-level system interacting with light (Figure~\ref{fig:figure2}a). If the light is polarized in the $\hat{x}$-direction, the susceptibility $\chi_{21}$ of this two-level system is given by 

\begin{equation}
    \label{Eq:two-level_susceptibility}
    \chi_{21}(\omega)=\frac{e^2d_{12}^2}{\hbar\varepsilon_0} \frac{T_2}{(\omega_0-\omega)T_2-i},
\end{equation}

\noindent where $\hbar\omega_0 = \hbar\omega_2 - \hbar\omega_1$ is the energy difference between the two levels corresponding to the ground state $\ket{1}$ and excited state $\ket{2}$, $d_{12}=\bra{1}\hat{x}\ket{2}$ is the dipole matrix element, $T_2$ is the mean dephasing time, $e$ is the electric charge, and $\varepsilon_0$ is the vacuum permittivity. The imaginary part of the susceptibility is a Lorentzian function centered at $\omega_0$, while the real part is non-zero for all $\omega<\omega_0$ and has a maximum at $\omega=\omega_0-1/T_2$. Extending this derivation to semiconductors (see Figure~\ref{fig:figure2}a) with a near-continuum of states in the valence and conduction bands requires summing up the contributions from each pair of states in the conduction and valence bands as

\begin{equation}
    \label{Eq:SC_susceptibility}
    \chi(\omega)=\frac{e^2T_2d_\textrm{cv}^2}{\hbar \varepsilon_0}\frac{2}{8\pi^3}\int_\textrm{BZ}\textrm{d}^3\textbf{k} \frac{\delta(E_\textrm{c}(\textbf{k})-E_\textrm{v}(\textbf{k})-\hbar\omega)}{(\omega_\textrm{cv}(\textbf{k})-\omega)T_2-i},
\end{equation}
\noindent where $d_\textrm{cv}$ is the dipole matrix element for the valence and conduction band states, and $\hbar\omega_\textrm{cv}=E_\textrm{c}(\textbf{k})-E_\textrm{v}(\textbf{k})$. The integrand counts the number of state pairs available across the conduction and valence bands for an optical transition of photon energy $\hbar \omega$ and weights it by the Lorentzian function. If the imaginary part of the Lorentzian is approximated by a delta function centered at $\omega$, the imaginary part of the integrand becomes the joint density-of-states (JDOS) given by

\begin{equation}
    \label{Eq:JDOS}
    \rho_\textrm{cv}(\omega)=\frac{2}{8\pi^3}\int_\textrm{BZ} \textrm{d}^3\textbf{k} \, \delta( E_\textrm{c}(\textbf{k})-E_\textrm{v}(\textbf{k})-\hbar\omega),
\end{equation}
\noindent where $E_\textrm{c}(\textbf{k})$ and $E_\textrm{v}(\textbf{k})$ are the conduction and valence band dispersion relationships of the semiconductor, and the integral is carried out over the entire Brillouin zone (BZ).

With the definition of JDOS, the imaginary and the real parts of permittivity, $\epsilon''(\omega)$ and $\epsilon'(\omega)$, are given by 

\begin{align}
\chi''(\omega)&=\frac{\pi e^2 d_\textrm{cv}^2}{2\hbar\varepsilon_0} \rho_\textrm{cv}(\omega), \nonumber \\
\chi'(\omega\ll E_\textrm{g}/\hbar)&\approx\frac{2}{\pi}\int_{E_\textrm{g}/\hbar}^{\infty} \frac{\chi''(\Omega)}{\Omega}d\Omega.
    \label{Eq:SC_susceptibility2}
\end{align}

 Equation~(\ref{Eq:SC_susceptibility2}) adequately describes the optical properties of many semiconductors. A large refractive index can arise from a large dipole matrix element $d_{cv}$ or a large JDOS $\rho_{cv}$. The dipole matrix element is directly related to the interband momentum matrix element or the Kane interband parameter. This parameter is nearly constant for many materials. Hence, the JDOS $\rho_{cv}$ is the primary parameter in determining the super-Mossian behavior of semiconductors.

To get a better insight into the JDOS, Eq.~(\ref{Eq:JDOS}) may be written in a different form. If we consider surfaces $S(E)$ of constant energy difference between the conduction and valence bands, $E(\textbf{k})=\hbar\omega(\textbf{k)=}E_\textrm{c}(\textbf{k})-E_\textrm{v}(\textbf{k})$, the JDOS may be expressed as~\cite{YuCardona2010} 

\begin{equation}
    \label{Eq:JDOS2}
    \rho_\textrm{cv}(\hbar\omega)=\frac{2}{8\pi^3}\iint_{S(\omega)} \frac{dS}{|\nabla_\textrm{k}(E_\textrm{c}-E_\textrm{v})|}.
\end{equation}

From Eq.~(\ref{Eq:JDOS2}), the JDOS grows significantly when the denominator of the integrand vanishes. Such points on $S$ are called joint critical points. Joint critical points can appear when the individual gradients, $\nabla_\textbf{k} E_\textrm{c}$ and $\nabla_\textbf{k} E_\textrm{v}$, vanish or when the conduction and valence bands track each other, i.e., $|\nabla_\textrm{k}(E_\textrm{c}-E_\textrm{v})| \approx 0$ (see Figure~\ref{fig:figure2}b). Both of these cases boost the JDOS. Most semiconductors meet the first requirement, i.e., the individual band gradients vanishing at the band extrema (corresponding to the direct band gap). At these points, the JDOS strengthens and contributes to the sub-bandgap refractive index. The second condition, i.e., bands tracking each other, happens not in all materials but in some, giving rise to their outstanding super-Mossian behavior. Both types of JDOS singularities are highlighted in the electronic band structure of silicon (Fig.~\ref{fig:figure2}c) and in the resulting refractive index (Fig.~\ref{fig:figure2}d). From the real part of the susceptibility in Eq.~(\ref{Eq:SC_susceptibility2}), the closer this type of joint critical point is to the band gap, the larger its contribution to the sub-bandgap index. Thus, semiconductors with conduction and valence bands tracking each other and their energy difference close to the optical band gap will likely be super-Mossian.

\section{Outstanding super-Mossian dielectrics}
\begin{figure*}
    \centering
    \includegraphics[width=17.2cm]{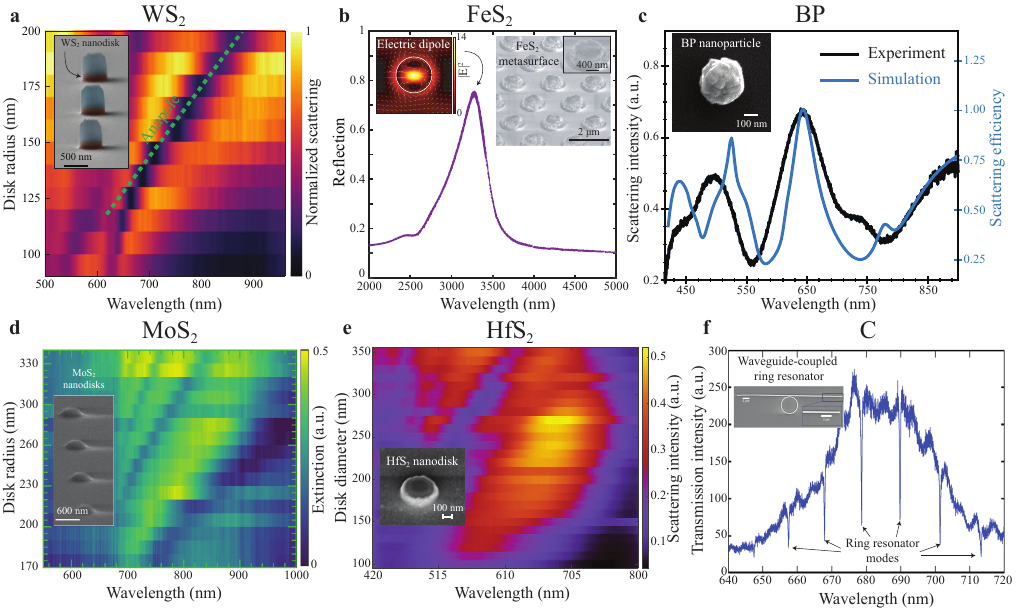}
    \caption{\textbf{Experimental demonstrations of super-Mossian materials in nanophotonics.}  \textbf{a} Dark-field scattering map of WS$_2$ nanodisks with varying disk radii. The anapole resonance is indicated by the green dashed line. Inset: scanning electron micrograph of the fabricated WS$_2$ nanodisks with a residual resist top layer. \textbf{b} Reflection spectrum of a FeS$_2$ metasurface supporting an electric dipole resonance. Inset: scanning electron micrograph of the fabricated metasurface. \textbf{c} Dark-field scattering (black) and simulated scattering efficiency (blue) of a BP nanoparticle supporting multiple Mie resonances. Inset: scanning electron micrograph of BP nanoparticle. \textbf{d} Extinction map of MoS$_2$ nanodisks with varying disk radii supporting multiple Mie resonances. Inset: scanning electron micrograph of fabricated MoS$_2$ nanodisks. \textbf{e} Dark-field scattering map of HfS$_2$ nanodisks with varying disk diameters supporting multiple Mie resonances. Inset: scanning electron micrograph of fabricated HfS$_2$ nanodisks. \textbf{f} Transmission spectrum of diamond waveguide coupled to a ring resonator showing dips due to excitation of ring resonator modes. Inset: scanning electron micrograph of waveguide-coupled ring resonator. Figures adapted with permission from: \textbf{a} Ref.~\cite{Verre2019}, Springer Nature; \textbf{b} Ref.~\cite{doiron2022super}, Wiley-VCH GmbH; \textbf{c-e} Refs.~\cite{Svendsen2022,Green2020,Zambrana-Puyalto2025} under a Creative Commons License \href{https://creativecommons.org/licenses/by/4.0/}{CC BY 4.0}; \textbf{f} Ref.~\cite{Hausmann2012}, American Chemical Society.}
    \label{fig:figure3}
\end{figure*}
The search for high-index dielectrics has accelerated in recent years, spanning both traditional semiconductors and emerging materials~\cite{sahoo2006achieving,baranov2017all,Gutierrez2018,Shim2020,doiron2022super,Svendsen2022,Yang2023,Zotev2025}. Early systematic studies established figures of merit for dielectric nanoresonators and showed that conventional IV and III–V semiconductors such as Si~\cite{Cao2010}, Ge~\cite{Cao2009}, GaSb, GeTe, and PbTe possess refractive indices that make them highly effective for photonic applications from the visible to the mid-infrared range~\cite{baranov2017all}. A subsequent screening of 338 monomer and binary semiconductors expanded this search and identified boron phosphide (BP) as a particularly promising dielectric for operation in the visible and ultraviolet regions~\cite{Svendsen2022}. The proximity of the optical band gap to the fundamental electronic band gap, together with a large density of states near the band edges, has been shown to give rise to super-Mossian behavior~\cite{khurgin2022expanding,doiron2022super}. Building on this understanding, a search through a comprehensive DFT database~\cite{petousis2017high} led to the identification and experimental demonstration of iron pyrite (FeS$_2$) as a near-infrared super-Mossian material~\cite{doiron2022super}.

Transition-metal dichalcogenides (TMDCs) in bulk multilayer form have recently emerged as a particularly promising class of super-Mossian layered dielectrics for nanophotonics~\cite{Munkhbat2023,Zotev2025}. Their high in-plane refractive indices and pronounced excitonic resonances make them attractive for realizing subwavelength light confinement and Mie-type resonances. A key experimental milestone was the demonstration of WS$_2$ nanodisks supporting distinct Mie and anapole resonances~\cite{Verre2019}, establishing WS$_2$ as a super-Mossian material for visible and near-infrared photonics~\cite{Zhang2020,Zotev2022,Zotev2023,Weber2023}. Subsequent studies have extended these concepts to other TMDCs, including MoS$_2$~\cite{Green2020,Ermolaev2021,Wang2025,Zotev2023}, WSe$_2$~\cite{Zotev2023}, and MoSe$_2$~\cite{Zotev2023}. Most recently, HfS$_2$ was identified through targeted screening of layered compounds and experimentally demonstrated as a super-Mossian dielectric~\cite{Zambrana-Puyalto2025}. 

Representative examples of experimentally realized super-Mossian materials are summarized in Figure~\ref{fig:figure3}, illustrating how this behavior spans diverse material classes. Figure~\ref{fig:figure3} highlights WS$_2$~\cite{Verre2019}, FeS$_2$~\cite{doiron2022super}, BP~\cite{Svendsen2022}, MoS$_2$~\cite{Green2020}, HfS$_2$~\cite{Zambrana-Puyalto2025}, and C (diamond)~\cite{Hausmann2012,Aharonovich2014}, each demonstrating resonant light–matter interaction consistent with their exceptionally high refractive indices. Beyond these well-studied systems, a wide range of materials are now being explored for potential super-Mossian characteristics, including GaS~\cite{Zotev2023,Gutierrez2024}, SiC~\cite{Lukin2020,Maruyama2025}, As$_2$S$_3$~\cite{Slavich2024}, GaP~\cite{Cambiasso2017,Shima2023}, AlN~\cite{Hu2020}, hBN~\cite{Caldwell2019,Kuhner2023}, GeS$_2$~\cite{Slavich2025},  HfSe$_2$~\cite{Zotev2023,Kowalski2025}, ReS$_2$~\cite{Shubnic2020,Mooshammer2022}, and SbSI~\cite{Starczewska2020}, among others. These studies demonstrate that super-Mossian behavior is not confined to a single chemical family but is a property that can manifest across traditional and emerging material systems.

\section{Optical materials discovery}
Rational materials design strategies based on first-principles computational screening of large material spaces have been widely applied to discover novel materials across diverse fields, including intermetallic alloys~\cite{curtarolo2012aflowlib}, thermoelectric materials~\cite{wang2011assessing},  (electro)catalysts~\cite{greeley2009alloys}, battery electrodes~\cite{hautier2011phosphates}, and inorganic solar cells~\cite{yu2012identification,kangsabanik2022indirect}. With a few exceptions discussed below, similar large-scale approaches have been pursued far less in the area of optical materials. This is likely because first-principles methods have historically played a limited role in photonics, where the governing physics often does not require an atomistic description, and because accurately treating excited states within first-principles frameworks remains challenging. 

The workhorse of first-principles materials science is density functional theory (DFT)~\cite{hohenberg1964inhomogeneous} and its time-dependent analogue, time-dependent density functional theory (TDDFT), which is used to describe response functions and excited states~\cite{runge1984density}. Although DFT is a ground-state theory, it yields an effective single-particle band structure similar to that obtained in Hartree--Fock theory. This DFT band structure is often used as an approximation to the true quasiparticle band structure. While the DFT band structure is often qualitatively accurate, the band gaps of semiconductors and insulators are systematically underestimated when standard semilocal exchange-correlation approximations are used~\cite{perdew1985density}. TDDFT inherits this shortcoming and, in addition, most common exchange-correlation kernels do not capture excitonic effects~\cite{onida2002electronic}. As a result, TDDFT with local exchange-correlation kernels generally offers little improvement over the random phase approximation, which neglects exchange-correlation effects in the dynamic response altogether. These shortcomings limit the quantitative accuracy of DFT- and TDDFT-based predictions of semiconductor optical properties, including their absorption spectra and refractive indices. 

Many-body perturbation theory provides a more accurate and predictive alternative to TDDFT, albeit at significantly higher computational cost. The gold standards for computing quasiparticle band structures and optical excitations are the GW self-energy method and the Bethe--Salpeter Equation (BSE), respectively~\cite{hybertsen1986electron,albrecht1998ab}. By explicitly accounting for self-energy effects and electron-hole interactions, the GW--BSE approach yields band gaps, exciton energies, and absorption onsets in close agreement with experiments, typically within 5-10\%, depending on the degree of self-consistency in the GW self-energy~\cite{van2006quasiparticle,shishkin2007self,huser2013quasiparticle}.

Due to the high computational cost of the BSE, it can be challenging to achieve convergence of the real part of the permittivity. Through the Kramers--Kronig relations, even the low-frequency response depends on contributions from high-frequency excitations. However, including a sufficient number of electron-hole transitions in the BSE two-particle Hamiltonian is often computationally infeasible. To adress this, an embedded BSE scheme (referred to as BSE+) was recently proposed to circumvent the slow convergence of the imaginary part of the permittivity by treating high-energy electron-hole transitions at the level of TDDFT~\cite{sondersted2024improved}. When tested on six common semiconductors, the BSE+ method was found to reduce the mean absolute percentage error of the low-frequency refractive index relative to experiments from 15.9\% (BSE) to 2.6\% (BSE+). For comparison, the more computationally efficient random phase approximation yielded an average error of 8.4\% for the same data set. An independent study compared the refractive index from TDDFT with experiments for 80 semiconductors and found an average error of 7\%.  

The TDDFT methodology has been used for high-throughput calculations of the dielectric properties of inorganic solids. In one study, the static dielectric tensor and refractive index of 1056 inorganic compounds has been calculated using density functional perturbation theory (equivalent to the static limit of TDDFT)~\cite{petousis2017high}. The materials were selected to have a DFT band gap above 0.1 eV and low formation energy, but they were not necessarily synthesized before.~\cite{petousis2017high}.
Using a similar computational approach, a subsequent study computed the static refractive index of 4040 previously synthesized semiconductors~\cite{naccarato2019searching} from the inorganic crystal structure database~\cite{belsky2002new}. Going beyond the static approximation, TDDFT has been applied to calculate the full frequency-dependent refractive index tensor of 338 experimentally known binary semiconductors~\cite{Svendsen2022}.  

Given the substantial effort required to synthesize and characterize new materials, reducing the uncertainty of computational predictions can yield significant time savings in the laboratory. For this reason, future optical materials discovery should move beyond the limited accuracy of TDDFT and the random phase approximation. In addition to the approximate treatment of the electronic response, all previous discovery studies have focused on the perfect crystalline material within the frozen-lattice approximation. Although this "ideal lattice" model provides an upper bound on the refractive index and a lower bound on optical losses, more realistic material models are needed to capture non-ideal samples. These include the effects of grain boundaries, point defects, impurities, and even amorphous structures. With deviations from a non-ideal crystal lattice comes the violation of momentum conservation leading to indirect transitions and possibly the formation of (localized) states inside the band gap. Lattice vibrations and electron-phonon coupling further contribute to indirect phonon-assisted transitions~\cite{noffsinger2012phonon,kangsabanik2022indirect}. Finally, non-radiative loss processes such as Shockley--Read--Hall recombination through defects states in the band gap and three-particle Auger recombination could also be important to consider to fully evaluate a material's potential for photonic applications.  

\section{Applications}
\begin{figure*}
    \centering
    \includegraphics[width=17cm]{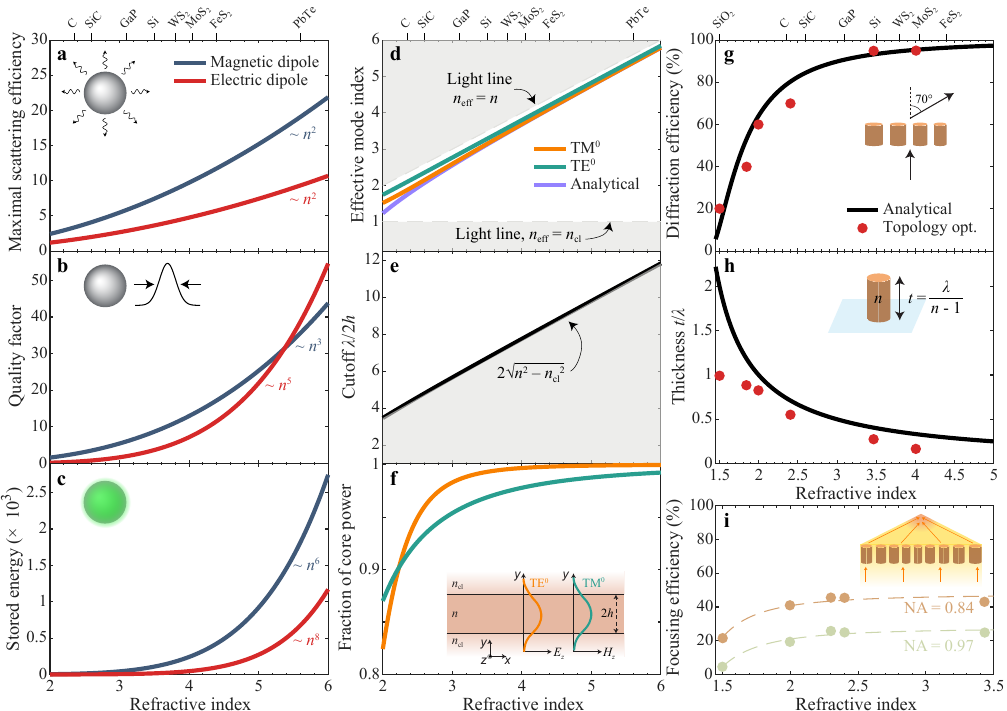}
    \caption{\textbf{Refractive index impact on nanophotonic devices.} \textbf{a}-\textbf{c} Maximal scattering efficiency, quality factor, and stored energy enhancement for the magnetic and electric dipole resonances supported by a spherical dielectric resonator as a function of refractive index, respectively. \textbf{d}-\textbf{f} Effective mode index, single-mode cut-off, and fraction of power guided in the waveguide core of a dielectric slab waveguide as a function of refractive index, respectively. The slab waveguide has thickness $2h$ and is surrounded by a cladding refractive index of $n_\textrm{cl}=1$. The wavenumber-thickness product is set to $k_0h=1$. The inset in \textbf{f} shows a schematic of the electric and magnetic field profiles of the TE$^0$ and TM$^0$ modes. \textbf{g} Diffraction efficiency (+1 order) of phase-discretized blazed grating as a function of refractive index. The grating diffracts normally incident light into an angle of $70^{\circ}$. \textbf{h} Minimum meta-atom thickness to achieve $2\pi$ phase shift as a function of refractive index. The red circles in \textbf{g}-\textbf{h} show the performance of topology-optimized dielectric meta-gratings from Ref.~\cite{Yang2017}. \textbf{i} Metalens focusing efficiency as a function of refractive index for different numerical apertures (NAs) from Ref.~\cite{Bayati2019}. The dashed lines are guides to the eye.}
    \label{fig:figure4}
\end{figure*}
\textbf{Nanoresonators.} Dielectric materials structured at the nanoscale, i.e., dielectric nanoresonators, are the fundamental building blocks for advanced optical devices. When such materials exhibit a high refractive index $n$, they can support optical resonances, known as Mie resonances, which can be both electric and magnetic in nature~\cite{Kuznetsov2016,Kruk2017}. To gain insight into their performance, we analyze the prototypical case of light scattering by a lossless dielectric spherical particle, as solved by Lorenz--Mie theory~\cite{Lorenz1890,Mie1908,Hulst1957,Bohren1983}. We show that the maximal scattering efficiency, the quality factor and the internal electromagnetic field enhancement are intrinsically dependent on the refractive index of the material. Specifically, a higher refractive index enhances these parameters, leading to superior optical performance. 

The scattering efficiency of a spherical particle with radius $R$ and refractive index $n$ illuminated by a plane wave with wavelength $\lambda$ is given as
\begin{equation} \label{eq:sigma_sca}
    \sigma_\textrm{sca} = \frac{2}{x^2} \sum_{j=1}^\infty (2j+1) \left( |a_j|^2 + |b_j|^2 \right), 
\end{equation}
where $x=2\pi R/\lambda$ is the size parameter, while $a_j$ and $b_j$ denote the electric-type and magnetic-type Mie coefficients of order $j$, respectively. In the limit of high-$n$, the maximal scattering efficiency of any Mie resonance is dictated by the refractive index of the spherical scatterer (see Supplementary~Note~1). For the magnetic and electric dipolar resonances, the maximal scattering efficiencies are given by
\begin{align} 
    \sigma_{\textrm{sca}}^\textrm{md} &= \frac{6}{\pi^2} n^2 \label{eq:sca_max_md} \\
    \sigma_{\textrm{sca}}^\textrm{ed} &= \frac{6}{4.493^2} n^2. \label{eq:sca_max_ed}
\end{align}
This analysis shows that the maximal scattering efficiency of any Mie resonance in a spherical scatterer scales with the refractive index squared (Fig.~\ref{fig:figure4}a). It is also worth noting that the resonant size parameter scales with the inverse of the refractive index (see Supplementary~Note~1), showing that, for a fixed resonance wavelength, smaller resonators may be employed. 
    
One of the most important parameters of nanoscale resonators are their quality ($Q$) factor as this characterizes the average photon lifetime in the cavity. Consequently, higher $Q$ factors result in sharper resonances and provide spectral sensitivity as well as enhanced light-matter interaction. Recent work~\cite{Zambrana-Puyalto2024} has analytically shown that the $Q$ factor of electric and magnetic dipole resonances in the high-$n$ limit ($n\gg$1) are given as
\begin{equation} \label{eq:Qj_mded}
\begin{aligned}
    Q^{\textrm{md}} & \approx  0.20 n^3, \\ 
    Q^{\textrm{ed}} & \approx  0.0070 n^5.
\end{aligned}
\end{equation}
Equation~(\ref{eq:Qj_mded}) shows that the $Q$ factor depends strongly on the refractive index (Fig.~\ref{fig:figure4}b). It is also worth noting that for refractive indices larger than approximately 5, the electric dipole mode has larger $Q$ factor than the magnetic dipole mode, and vice versa (Fig.~\ref{fig:figure4}b). A similar strong dependence on the refractive index has been identified in subwavelength dielectric supercavity resonators~\cite{Rybin2017}.

Optical processes such as nonlinear interactions, Raman scattering, and photoluminescence can be significantly enhanced by the local amplification of electromagnetic fields in nanoscale resonators. To quantify this enhancement in spherical Mie resonators, we use the internal stored energy as a measure of the local field intensity. The stored energy is directly related to the $Q$ factor of the resonator. In Supplementary~Note~2, we derive an expression for its enhancement as a function of the refractive index. Specifically, the stored energy enhancement, relative to free space, for magnetic and electric dipole resonances in the high-$n$ limit is given as
\begin{equation} \label{eq:etaj}
\begin{aligned}
    \eta^{\textrm{md}} & \approx  0.058 n^6, \\ 
    \eta^{\textrm{ed}} & \approx  0.0007 n^8.
\end{aligned}
\end{equation}
Notably, the stored energy enhancement exhibits an additional scaling factor of $n^3$ compared to the $Q$ factor (Fig.~\ref{fig:figure4}c). While these relations describing the performance of dielectric resonators have centered around the spherical geometry, similar performance boosts due to increasing the refractive can be expected for resonators with other geometries.

\textbf{Waveguides.} Guiding and confining light in near-lossless dielectric materials is fundamental to integrated photonics, where optical signals are manipulated on compact chips. Light is typically routed through strip waveguides, i.e., rectangular structures made from a high-index dielectric surrounded by a lower-index cladding material. The contrast in refractive index between core and cladding is what enables confinement, and thus plays a critical role in determining device size, efficiency, and overall integration density.

To examine the impact of refractive index on waveguides, we consider a simplified model: a symmetric dielectric slab waveguide of thickness $2h$ with core refractive index $n$ surrounded by a cladding of refractive index $n_\textrm{cl}$. This geometry allows for analytical treatment of guided modes and highlights how increasing the refractive index enables tighter mode localization and thereby reduced device footprint. In the following, we examine three important parameters of the dielectric slab waveguide as a function of its refractive index; the effective index, the cut-off frequency for single-mode operation, and the fraction of power in the waveguide core.

The dielectric slab waveguide supports both transverse-electric (TE) and transverse-magnetic (TM) modes, where the electric and magnetic field have only in-plane components, respectively~\cite{Balanis2012}. The dispersion relation for the fundamental TE$^0$ and TM$^0$ modes are given by
\begin{equation}\label{eq:slabdispersion}
    k_0h\sqrt{n^2-n_\textrm{eff}^2} = \arctan\left( \tau \sqrt{\frac{n_\textrm{eff}^2-n_\textrm{cl}^2}{n^2-n_\textrm{eff}^2}} \right),
\end{equation}
where $k_0=2\pi/\lambda$ is the free-space wavenumber, $n_\textrm{eff}$ denotes the effective mode index, and $\tau=(n/n_\textrm{cl})^2$ for TM and $\tau=1$ for TE. In the limit of high-$n$, the effective index is close to the refractive index of the core ($n_\textrm{eff}\approx n)$ and we can perform the approximation $\lim_{x\rightarrow\infty} \arctan(x)=\pi/2$ to obtain a simplified expression for the effective mode index
\begin{equation}\label{eq:effectiveindex}
    n_\textrm{eff} \simeq \sqrt{n^2 - \left(\frac{\pi}{2hk_0}\right)^2}.
\end{equation}
Equation~(\ref{eq:effectiveindex}) applies for both of the fundamental TE$^0$ and TM$^0$ modes. The analytical relation in Eq.~(\ref{eq:effectiveindex}) shows good agreement with the exact numerical solution of Eq.~(\ref{eq:slabdispersion}) (see Fig.~\ref{fig:figure4}d) and highlights that the effective index increases nearly linearly with increasing refractive index of the dielectric waveguide. 

We now turn to the concept of cut-off frequency in slab waveguides as a means to evaluate when the structure supports only the fundamental modes. Unlike strip waveguides, which confine light in both vertical and lateral directions and therefore exhibit cut-off frequencies even for the fundamental modes, the slab waveguide confines light in only one dimension. As a result, the fundamental modes have no cut-off frequency in the slab geometry. To enable a meaningful comparison with strip waveguides, we instead consider the frequency below which the slab supports only the fundamental TE$^0$ and TM$^0$ modes. This serves as an analogue to the cut-off behavior observed in strip waveguides. The wavelength-to-thickness ratio where the slab waveguide only supports the fundamental modes is given by~\cite{Balanis2012}
\begin{equation}\label{eq:cutoff}
    \frac{\lambda}{2h} > 2\sqrt{n^2-n_\textrm{cl}^2}.
\end{equation}
For a fixed wavelength, a waveguide core with higher refractive index enables the use of a thinner and more compact waveguide (see Fig.~\ref{fig:figure4}e). Ultimately, this would enable increased integration density in integrated photonic circuits.

Finally, we examine the fraction of power guided in the core of a slab waveguide. This is evaluated as the modal power density, given by the Poynting vector, integrated in the core of the waveguide over the total modal power (i.e., integrated over all of space)~\cite{Snyder1984}, and is given by 
\begin{equation} \label{eq:powercore}
    \eta_\textrm{w} = 1 - \frac{n^2-n_\textrm{eff}^2}{n^2-n_\textrm{cl}^2}\left(1+fk_0h\sqrt{n_\textrm{eff}^2-n_\textrm{cl}^2}\right)^{-1},
\end{equation}
where $f=1$ for TE modes and $f=n_\textrm{eff}^2(n^2+n_\textrm{cl}^2)/n^2n_\textrm{cl}^2-1$ for TM modes. Figure~\ref{fig:figure4}f shows that a significant increase in the power guided in the core can be realized with higher core refractive index, which may be leveraged for constructing waveguides with smaller radii of curvature.

\textbf{Metasurfaces.} Metasurfaces, including meta-gratings and metalenses, achieve control over an incident beam of light by precisely structuring phase profiles at subwavelength scales.  Particularly, for metalenses with high numerical aperture, rays at the lens periphery require substantial angular deflection, analogous to meta-gratings diffracting at large angles. To quantitatively explore the role of the refractive index in meta-optical components, we approximate the metasurface locally as a phase-discretized blazed grating.

A blazed grating exhibits a linear sawtooth phase profile that repeats every period $\Lambda$. The diffraction angle $\theta$ for normal incidence illumination is given by the grating equation $\sin\theta = \lambda/\Lambda$. When fabricating metasurfaces, the continuous linear phase ramp is discretized into $N$  levels. According to Fourier optics theory~\cite{Goodman2017}, the first-order diffraction efficiency $\eta_\textrm{d}$ of a discretized blazed grating follows the relationship
\begin{equation} \label{eq:diffeffN}
\eta_\textrm{d} = \mathrm{sinc}^2\left(\frac{1}{N}\right).
\end{equation}
To connect the refractive index $n$ to the discretization number $N$, we note that achieving full $2\pi$ phase coverage with meta-atoms of refractive index $n$ and height $t$ requires~\cite{Moon2022}
\begin{equation}\label{eq:metaatomheight}
t \approx \frac{\lambda}{n - 1}.
\end{equation}
Fabrication constraints typically limit the lateral size (unit cell period) of a meta-atom $p$ to approximately its height scaled by a structural factor $\alpha \approx 1$, thus $p\approx \alpha t$. Given the grating period $\Lambda$, the maximum achievable number of discrete phase levels is then $N_{\mathrm{max}}=\Lambda/p = (n-1)/(\alpha \sin\theta)$, where we have used the grating equation. Substituting this result into Eq.~(\ref{eq:diffeffN}), we obtain a relationship between the refractive index and the diffraction efficiency
\begin{equation} \label{eq:diffefffinal}
\eta_\textrm{d}= \mathrm{sinc}^2\left(\alpha\frac{ \sin\theta}{n - 1}\right).
\end{equation}
Figure~\ref{fig:figure4}g illustrates the case of diffracting normally incident light to an angle of $70^{\circ}$, using Eq.~(\ref{eq:diffefffinal}) with $\alpha=1.2$. We compare our analytical relation to topology-optimized meta-gratings (red circles)~\cite{Yang2017}, and find excellent agreement, demonstrating that increasing the refractive index of the metasurface material substantially enhances diffraction efficiency by enabling finer phase discretization. A higher refractive index also allows for thinner meta-atoms, which may reduce fabrication complexity~(Fig.~\ref{fig:figure4}h). 

Shifting from metagratings to metalenses, Ref.~\cite{Bayati2019} studied how refractive index affects metalens focusing efficiency across a range of numerical apertures. As shown in Fig.~\ref{fig:figure4}i, their results similarly indicate improved performance with increasing index. Beyond these applications, increasing the refractive index can also improve metalens achromaticity~\cite{Presutti2020} and reduce overall optical device thickness~\cite{Miller2023}. Together, these findings underscore the critical importance of discovering and leveraging new high-index materials in metasurface applications.

\section{Outlook and conclusion}
Super-Mossian materials have the potential to enhance performance across virtually all areas of nanophotonics, since the efficiency of nanoresonators, waveguides, and metasurfaces scales directly with the refractive index. A higher refractive index enables stronger optical confinement, higher quality factor resonances, and improved light–matter interaction. This review has surveyed the current landscape of high-index materials, ranging from conventional semiconductors to emerging materials. From this overview, it is evident that super-Mossian behavior is not confined to a single chemical family but can manifest across diverse classes of materials.

The physical origin of super-Mossian behavior lies in a large joint density of states near the band edges, which enhances absorption just above the band gap. Such behavior typically arises from flat conduction or valence bands, or from situations where these bands closely track each other in momentum space. For a material to exhibit strong super-Mossian characteristics, the absorption onset associated with this large joint density of states should occur as close as possible to the fundamental band gap energy. Identifying such electronic structures is, however, nontrivial, and first-principles approaches can play a crucial role in accelerating this search.

Computational screening using density functional theory and related first-principles methods can substantially reduce the time and cost associated with experimental materials discovery. Although some studies have already explored this direction, the available chemical space remains far from exhausted. Machine-learning-assisted screening methods could further accelerate discovery, while generative models might even propose previously unknown compounds with tailored optical properties. To translate these computational predictions into real-world materials, close collaboration between theorists and experimentalists will be essential. Given that super-Mossian behavior occurs across many material classes, interdisciplinary engagement, from solid-state chemists to materials scientists in crystal growth and thin-film synthesis, will be key to advancing the field.

From an applications standpoint, practical progress will depend on the ability to synthesize high-quality thin films on low-loss substrates and to develop lithographic and etching protocols that yield reproducible nanostructures. No single material will suit all photonic applications as optimal performance depends strongly on the operational wavelength. Infrared devices, for instance, can benefit from lower-band-gap materials with correspondingly higher refractive indices, while visible and ultraviolet applications require wide-band-gap dielectrics that maintain high index without significant absorption.

The exploration of super-Mossian dielectrics is still in its early stages but holds great promise. By bridging first-principles computation, materials synthesis, and nanophotonic design, the next generation of high-index dielectrics could provide a substantial performance boost across photonic technologies and help translate laboratory demonstrations into new optical technology.

%\appendix
%\section{}

%\begin{acknowledgments}
%\end{acknowledgments}

\def\bibsection{\section*{References}}
\bibliography{references}

\end{document}